# Possible Current-Induced Phenomena and Domain Control in an Organic Dirac Fermion System with Weak Charge Ordering

Toshihito Osada* and Andhika Kiswandhi

*Institute for Solid State Physics, University of Tokyo,*

*5-1-5 Kashiwanoha, Kashiwa, Chiba 277-8581, Japan.*

We show that when the electron and hole densities are unbalanced, observable nonlinear anomalous Hall effect and current-induced orbital magnetization appear at zero magnetic field in the weak charge ordering (CO) state of an organic two-dimensional Dirac fermion system, $\alpha$-(BEDT-TTF)$_2$I$_3$. These current-induced phenomena are caused by a finite Berry curvature dipole resulting from inversion symmetry breaking and Dirac cone tilting. In the actual system, however, these effects are canceled out between different types of inversion asymmetric CO domains. To avoid the cancellation, we propose a new experimental method to realize the selective formation of a single type of domain using the current-induced magnetization.

A layered organic conductor, $\alpha$-(BEDT-TTF)$_2$I$_3$, is known as a two-dimensional (2D) massless Dirac fermion system with a pair of Dirac cones [1]. At ambient pressure, it undergoes a metal-insulator transition into an insulating phase due to charge ordering (CO) at $T_{CO}$ = 135 K. This CO phase is suppressed by applying pressure and vanishes at approximately $P_c$ = 1.3 GPa. Above $P_c$, a metallic massless Dirac fermion state is realized down to low temperatures. The Dirac fermion state was originally pointed out theoretically [2] and later confirmed by various experiments, such as magnetotransport [3, 4], specific heat [5], and NMR [6]. In the massless Dirac fermion state, the band dispersion around the Fermi level $E_F$ is identified as a pair of tilted and anisotropic Dirac cones with Dirac points located at general points, $\mathbf{k}_0$ and $-\mathbf{k}_0$, in the $\mathbf{k}$ space, which are referred to as the $\mathbf{k}_0$ and the $-\mathbf{k}_0$ valleys, respectively. They are energetically degenerated forming a Kramers pair due to time reversal symmetry, and each of them is spin-degenerate. In this study, we consider the "weak" CO state just below $P_c$, where the CO transition temperature $T_{CO}$ is largely suppressed. The electronic structure around $E_F$ is regarded as a massive Dirac fermion system, with a small CO gap opening at the Dirac points in the two valleys due to the breaking of inversion symmetry [7, 8]. In the weak CO state, the system generally consists of two types of CO domains, reflecting the breaking of inversion symmetry.

Recently, current-induced phenomena have been studied in inversion-asymmetric conductors. Current-induced orbital magnetization (orbital Edelstein effect), which is a kind of electromagnetic effect, was observed in strained monolayer MoS$_2$ [9, 10] and discussed in crystals with helical structure [11]. The nonlinear anomalous Hall effect (AHE) (current-induced AHE) was reported in thin films of a Weyl semimetal WTe$_2$ [12, 13], and was also discussed in a chiral crystal of Te [14]. These materials have finite Berry curvature, but a sum of the Berry curvature over all occupied states is zero at the equilibrium state, implying that they are topologically trivial conductors. However, when an electric current is applied to these conductors, the electron distribution is shifted in the **k** space, resulting in a finite sum of Berry curvature over occupied states. Therefore, the phenomena originating from the Berry curvature are induced in their current carrying state.

This study highlights the possibility of observing the current-induced phenomena, namely, nonlinear AHE and current-induced orbital magnetization in the weak CO state of an organic Dirac fermion system $\alpha$-(BEDT-TTF)$_2$I$_3$, when a finite imbalance exists between electron and hole densities. Furthermore, an experimental method is proposed to align the inversion-asymmetric CO domains in order to observe the current-induced phenomena in an actual system consisting of multiple CO domains.

First, we consider a massive Dirac fermion system with no inversion symmetry as a model for a single CO domain system. We employ the following tilted Weyl equation for the $\mathbf{k}_0$ valley of $\alpha$-(BEDT-TTF)$_2$I$_3$ [15, 16].

$$H(\mathbf{k}) = \hbar v_\mathrm{F}(k_x \sigma_x + k_y \sigma_y) + \hbar v_0 k_x \sigma_0 + \Delta \sigma_z. \tag{1}$$

Here, $\mathbf{k} = (k_x, k_y)$ is the 2D wave number measured from $\mathbf{k}_0$. The matrices $\sigma_x$, $\sigma_y$, and $\sigma_z$ are Pauli matrices and $\sigma_0$ is a $2\times 2$ unit matrix. Their bases correspond to the sublattice states. The $x$-axis is taken to be along the tilting direction of the Dirac cone, and $v_0$ indicates the amplitude of the tilting. In addition, we introduced the mass parameter $\Delta$, which breaks the inversion symmetry and opens the CO energy gap. The energy dispersion and the normal component of Berry curvature of the conduction (+) and the valence (−) bands are easily obtained as follows.

$$E_{\pm}(\mathbf{k}) = \hbar v_0 k_x \pm \sqrt{(\hbar v_F)^2 k_x^2 + (\hbar v_F)^2 k_y^2 + \Delta^2} \, , \tag{2}$$

$$[\mathbf{\Omega}_{\pm}(\mathbf{k})]_z = \mp (\hbar v_F)^2 \Delta \Big/ \left\{ 2\sqrt{(\hbar v_F)^2 k_x^2 + (\hbar v_F)^2 k_y^2 + \Delta^2}^{\,3} \right\} . \tag{3}$$

Note that the Berry curvature does not depend on the tilting.

For the $-\mathbf{k}_0$ valley, the effective Hamiltonian is given by the time reversal of (1): $H(\mathbf{k}) = \hbar v_F (-k_x \sigma_x - k_y \sigma_y^*) - \hbar v_0 k_x \sigma_0 + \Delta \sigma_z$, where $\mathbf{k}$ is measured from $-\mathbf{k}_0$. Note that $(\sigma_x, \sigma_y, \sigma_z)$ corresponds not to real spin, but sublattice pseudo-spin. The $-\mathbf{k}_0$ valley has a dispersion with opposite tilting slope $(-v_0)$ and a Berry curvature with an opposite sign. The dispersion and the Berry curvature of the $\mathbf{k}_0$ and $-\mathbf{k}_0$ valleys are schematically represented in Fig. 1, where the Berry curvature of the conduction band $[\mathbf{\Omega}_+(\mathbf{k})]_z$ is illustrated by contours.

An electron with a Berry curvature $\mathbf{\Omega}_{\pm}(\mathbf{k})$ obtains an anomalous velocity $\mathbf{v}_{\pm}^{(a)}(\mathbf{k}) = (e/\hbar) \mathbf{E} \times \mathbf{\Omega}_{\pm}(\mathbf{k})$ under an in-plane electric field $\mathbf{E}$ [17]. In the equilibrium state, the sum of the total electron velocity $\mathbf{v}_{\pm}(\mathbf{k}) = (1/\hbar) \partial E_{\pm}(\mathbf{k})/\partial \mathbf{k} + \mathbf{v}_{\pm}^{(a)}(\mathbf{k})$ over the occupied states vanishes because of the cancellation of the two valleys with opposite signs of $[\mathbf{\Omega}_{\pm}(\mathbf{k})]_z$, thus AHE never appears. Similarly, the electron obtains an orbital magnetic moment $\mathbf{m}_{\pm}(\mathbf{k}) = (e/\hbar)[\{E_{\pm}(\mathbf{k}) - E_{\mp}(\mathbf{k})\}/2] \mathbf{\Omega}_{\pm}(\mathbf{k})$ due to the self-rotation of the wave packet in the two-band system [17]. Here, we use the approximation formula $\mathbf{m}_{\pm}(\mathbf{k}) \simeq \pm (e/\hbar)(E_{gap}/2) \mathbf{\Omega}_{\pm}(\mathbf{k})$, where $E_{gap} \simeq 2|\Delta|$ denotes the average energy gap between the conduction and the valence bands. In the equilibrium state, the sum of $\mathbf{m}_{\pm}(\mathbf{k})$ over occupied states also vanishes because of the cancellation of the two valleys, resulting in no orbital magnetization. These results are expected from the time reversal symmetry.

Subsequently, we consider the non-equilibrium current carrying state, wherein the in-plane electric field **E** is applied to the conducting system and a stationary electric current is flowing. According to the Boltzmann transport theory, the distribution function is shifted by $\Delta f_{\pm}(\mathbf{k}) = (-\partial f_{\pm}^0(\mathbf{k})/\partial E)\tau \mathbf{v}_{\pm}(\mathbf{k}) \cdot (-e)\mathbf{E}$ in response to **E**. Here, $f_{\pm}^0(\mathbf{k}) = 1/[\exp\{(E_{\pm}(\mathbf{k}) - \mu)/k_B T\} + 1]$ is the equilibrium distribution with a chemical potential $\mu$, and $\tau$ denotes a constant relaxation time. In the **k** space, the occupied states move by $-(e\tau/\hbar)\mathbf{E}$, so that the anomalous Hall current and the orbital magnetization summed up over occupied states may become finite. Suppose that the electric field **E** is applied along the $x$-axis, which is the tilting axis of the Dirac cone in the present model. We also assume a small imbalance between electron and hole densities ($n - p > 0$), in other words, a positive Fermi energy at zero temperature. As illustrated in Fig. 1, the colored occupied region is shifted from the Fermi surface to the $-x$ direction in both valleys. Since the Berry curvature is unevenly distributed in the occupied region due to the tilting, the negative total Berry curvature $[\mathbf{\Omega}_+(\mathbf{k})]_z$ of the occupied states in the $\mathbf{k}_0$ valley decreases due to the shift, whereas the positive total Berry curvature of the $-\mathbf{k}_0$ valley increases. Therefore, the contribution from both valleys does not cancel completely under **E**, resulting in current-induced AHE and magnetization. The former is called nonlinear AHE because it is a second-order response to the electric field. This can be intuitively understood as a breaking of the balance of the average anomalous velocity, as illustrated in Fig. 1(c). The nonlinear AHE can also be interpreted as Hall effect due to the current-induced magnetization.

The magnitude of these current-induced effects is represented by the following

Berry curvature dipole (BCD) in the equilibrium state [18].

$$\Lambda_\pm^z \equiv \frac{1}{(2\pi)^3}\iiint \frac{\partial[\Omega_\pm(\mathbf{k})]_z}{\partial \mathbf{k}} f_\pm^0(\mathbf{k})d\mathbf{k}^3 \ . \tag{4}$$

We evaluated the BCD of a 2D tilted massive Dirac fermion system as the following. The $y$-component of the BCD $[\Lambda_\pm^z]_y$ is always zero, because the integrand of (4) is an odd function of $k_y$. As for the $x$-component $[\Lambda_\pm^z]_x$, the $\mathbf{k}_0$ and $-\mathbf{k}_0$ valleys have equal contributions. Figure 2(a) shows the chemical potential dependence of the BCDs of the conduction band $[\Lambda_+^z]_x$ and the valence band $[\Lambda_-^z]_x$ for several tilting cases at $T$=0. Here, $[\Lambda_\pm^z]_x$ is normalized by $3\hbar v_F/(2\pi^2 c\Delta)$ including spin degeneracy, where $c$ is the interlayer distance. The chemical potential $\mu$, which is equal to the Fermi energy at $T$=0, is normalized by $|\Delta|$ in the figure. We find that $[\Lambda_\pm^z]_x = 0$ in the case of no Dirac cone tilting ($v_0/v_F = 0$), and $[\Lambda_\pm^z]_x$ grows larger as the tilting amplitude increases. The sign of $[\Lambda_\pm^z]_x$ is reversed when $v_0/v_F$ is negative. In the case of $|v_0/v_F| > 1$, the system becomes a type-II Dirac system with hyperbolic Fermi lines. The flat part around $\mu/|\Delta| = 0$ corresponds to the energy gap between the conduction and the valence bands.

Figure 2(b) presents the chemical potential dependence of the sum and the difference of the conduction and the valence band BCDs, $[\Lambda_+^z \pm \Lambda_-^z]_x$, for several temperatures at a fixed tilting, $v_0/v_F$ =0.8. We can see that at finite temperatures, $[\Lambda_+^z \pm \Lambda_-^z]_x$ has a finite value even when $\mu$ is located in the energy gap, because thermally excited carriers in the conduction and the valence bands contribute to the BCD. This means that the BCD can appear even in the insulating state. In particular, $[\Lambda_+^z - \Lambda_-^z]_x$ remains finite even at $\mu=0$, where $[\Lambda_+^z + \Lambda_-^z]_x$ becomes zero.

The second-order current response against the electric field **E** is written as $j_\alpha^{(2)} = \chi_{\alpha\beta\beta} E_\beta^2$, using Einstein's notation. The response coefficient $\chi_{\alpha\beta\beta}$ (nonlinear Hall conductivity) relates to the BCD by $\chi_{\alpha\beta\beta} = (e^3\tau/\hbar^2)\varepsilon_{\alpha\beta z}[\Lambda_+^z + \Lambda_-^z]_\beta$, where $\varepsilon_{\alpha\beta\gamma}$ is the Levi-Civita symbol [18]. Because $[\Lambda_+^z + \Lambda_-^z]_y = 0$, the nonlinear AHE in the present system is represented as

$$\mathbf{j}^{(2)} = \left(\chi_{yxx} E_x^2\right)\mathbf{n}_y = \left\{-(e^3\tau/\hbar^2)[\Lambda_+^z + \Lambda_-^z]_x |\mathbf{E}|^2 \cos^2\theta\right\}\mathbf{n}_y, \qquad (5)$$

where $\mathbf{n}_y$ is the unit vector in the y-direction, and $\theta$ denotes the angle between **E** and the tilting axis (x-axis). Note that the nonlinear Hall current density $\mathbf{j}^{(2)}$ is always perpendicular to the tilting axis regardless of the electric field direction. The value of $j_y^{(2)}$ depends only on the tilting axis component $E_x$. In addition, note that the sign of $j_y^{(2)}$ never changes even if the electric field **E** is reversed, because it is a second-order response. The dependence of $j_y^{(2)}$ on the orientation of **E** is schematically illustrated in Fig. 3(a).

The current-induced magnetization is represented by $M_\alpha = \alpha_{\alpha\beta}^{\text{ME}} E_\beta$, where the magnetoelectric coefficient $\alpha_{\alpha\beta}^{\text{ME}}$ relates to the difference between the conduction and the valence band BCDs by $\alpha_{\alpha\beta}^{\text{ME}} \simeq -(e^2\tau E_{\text{gap}}/2\hbar^2)[\Lambda_+^\alpha - \Lambda_-^\alpha]_\beta$ [10]. The minus symbol before $\Lambda_-^\alpha$ originates from the difference of the sign between $\mathbf{m}_-(\mathbf{k})$ and $\mathbf{\Omega}_-(\mathbf{k})$. Because $[\Lambda_+^z - \Lambda_-^z]_y = 0$, the current-induced magnetization is given by

$$\mathbf{M} = \left(\alpha_{zx}^{\text{ME}} E_x\right)\mathbf{n}_z \simeq \left\{-(e^2\tau|\Delta|/\hbar^2)[\Lambda_+^z - \Lambda_-^z]_x |\mathbf{E}|\cos\theta\right\}\mathbf{n}_z, \qquad (6)$$

where $\mathbf{n}_z$ is the unit vector in the z direction. The orbital magnetization **M** is always normal to the 2D plane. The value of **M** depends only on the tilting axis component $E_x$, and changes its sign when the electric field is reversed, as depicted schematically in Fig. 3(b). Note that the magnetization is not scaled by $|\Delta|$, but it only depends on the sign of $\Delta$, because the factor $|\Delta|$ in (6) is canceled by the denominator of the normalizing factor $3\hbar v_F/(2\pi^2 c\Delta)$ of $[\Lambda_+^z - \Lambda_-^z]_x$.

In $\alpha$-(BEDT-TTF)$_2$I$_3$, the band parameters were estimated as $v_F = 1.0 \times 10^5$ m/s and $v_0 = 0.8 \times 10^5$ m/s from the comparison with the first principles calculation [16]. The interlayer distance was reported as $c = 1.75$ nm [19]. Using these values, we can quantitatively estimate the response coefficients of nonlinear AHE ($\chi_{yxx} = j_y^{(2)}/E_x^2$) and current-induced magnetization ($\alpha_{zx}^{ME} = M_z/E_x$) in $\alpha$-(BEDT-TTF)$_2$I$_3$. Figure 3(c) shows the carrier density dependence of $\chi_{yxx}$ and $\alpha_{zx}^{ME}$, which are normalized by the mass parameter $\Delta$ and the scattering relaxation time $\tau$. The carrier density imbalance $n-p$ is the difference in the densities between thermally excited electrons and holes.

Actual $\alpha$-(BEDT-TTF)$_2$I$_3$ crystals are slightly electron-doped, possibly due to the partial lack of I$_3^-$ ions. They show a finite Hall effect, suggesting an uncompensated carrier density $n-p$ in the order of $10^{15} \sim 10^{16}$ cm$^{-3}$ at low temperatures [1]. The mass parameter is estimated from $|\Delta| \sim k_B T_{CO}$. In a previous work, the scattering broadening of the $n=0$ Landau level was experimentally estimated to be 3 K [20], which corresponds to $\tau \sim 2.5$ ps if the scattering time does not change under the magnetic field. Assuming the realistic parameters, that is, $n-p = 10^{16}$ cm$^{-3}$, $|\Delta| = 1$ meV ($\sim 10$ K), and $\tau = 1$ ps, we can obtain $\chi_{yxx} = j_y^{(2)}/E_x^2 = 0.15$ A/V$^2$ and $\alpha_{zx}^{ME} = M_z/E_x = 0.15$ mA/V at $T = 0$ from Fig. 3(c).

These values are in the observable range.

So far, we have discussed the single-domain system. In the actual $\alpha$-(BEDT-TTF)$_2$I$_3$ crystal, however, multiple CO domains appear after the CO phase transition. In the multi-domain system, the nonlinear AHE and the current-induced magnetization can hardly be observed because of the cancellation between two types of inversion-asymmetric domains.

To observe the current-induced phenomena in the weak CO state, we have to realize the selective formation of one of the two types of domains. For this purpose, we propose a new experimental method utilizing the current-induced magnetization, which is referred as the current-field-cooling. As schematically depicted in Fig. 4, we slowly cool the sample under an applied DC electric current and a normal magnetic field, particularly around the CO transition. The electric current direction parallel to the tilting direction of Dirac cones is most effective. When CO domains are formed around the transition, each domain has a current-induced magnetization either parallel or antiparallel to the normal of the 2D plane. Therefore, it feels a potential $U$ under the magnetic field depending on the magnetization direction. The potential is denoted by $U = -V_{\text{domain}}(\mathbf{M}\cdot\mathbf{B}) = -V_{\text{domain}}\alpha_{zx}^{\text{ME}}E_xB_z$, where $V_{\text{domain}}$ is the domain volume. When $|U|$ is larger than $k_BT_{\text{CO}}$, one type of the domain is selectively formed by the DC current and the magnetic field during the fluctuating domain formation. This current-field-cooling mechanism likely works in the actual system, since the potential can be estimated to be $|U| \sim 400$ K $> k_BT_{\text{CO}}$ for realistic parameters, $V_{\text{domain}} = 1$ $\mu$m$^3$, $\alpha_{zx}^{\text{ME}} = 0.6$ mA/V ($k_BT/|\Delta|=1$ is assumed), $E_x = 1$ mV/mm, and $B_z = 10$ T.

We might use the current-induced phenomena to investigate the unidentified insulating state of $\alpha$-(BEDT-TTF)$_2$I$_3$. The inset of Fig. 4 displays the temperature dependence of interlayer resistance of $\alpha$-(BEDT-TTF)$_2$I$_3$ at several pressures (after Mori *et al*. [21]). At $P$ = 1.2 GPa, just below the critical pressure, the system undergoes a transition to the weak CO state at $T_{CO}$. It is visible that the system exhibits another insulating behavior at low temperatures below the dotted arrow [22, 23]. The excitonic instability [24] and the topological insulator state [25] were proposed as the origin of this unidentified insulating state. Note that inversion symmetry remains in both cases. However, if the current-induced phenomena are observed in this unidentified insulating state obtained by the current-field-cooling, a gapped state with broken inversion symmetry is strongly suggested.

In conclusion, we studied the possible nonlinear AHE and current-induced orbital magnetization at zero magnetic field in the weak CO state of an organic conductor $\alpha$-(BEDT-TTF)$_2$I$_3$, which is a 2D massive Dirac fermion system with Dirac cone tilting and gap opening due to inversion symmetry breaking. This is an ideal system with a finite BCD, and we demonstrated that a single CO domain system exhibits observable nonlinear AHE and current-induced magnetization in the current carrying state. To avoid the cancellation of current-induced effects between different types of domains, we proposed a current-field-cooling method to enhance the selective formation of a single type of domain.

**Acknowledgements**

The authors thank Dr. T. Taen, Dr. K. Uchida, Ms. A. Mori, Mr. K. Yoshimura, and Dr. M. Sato for their valuable input. This work was supported by JSPS KAKENHI Grant No. JP20H01860.

# References

*corresponding author, osada@issp.u-tokyo.ac.jp

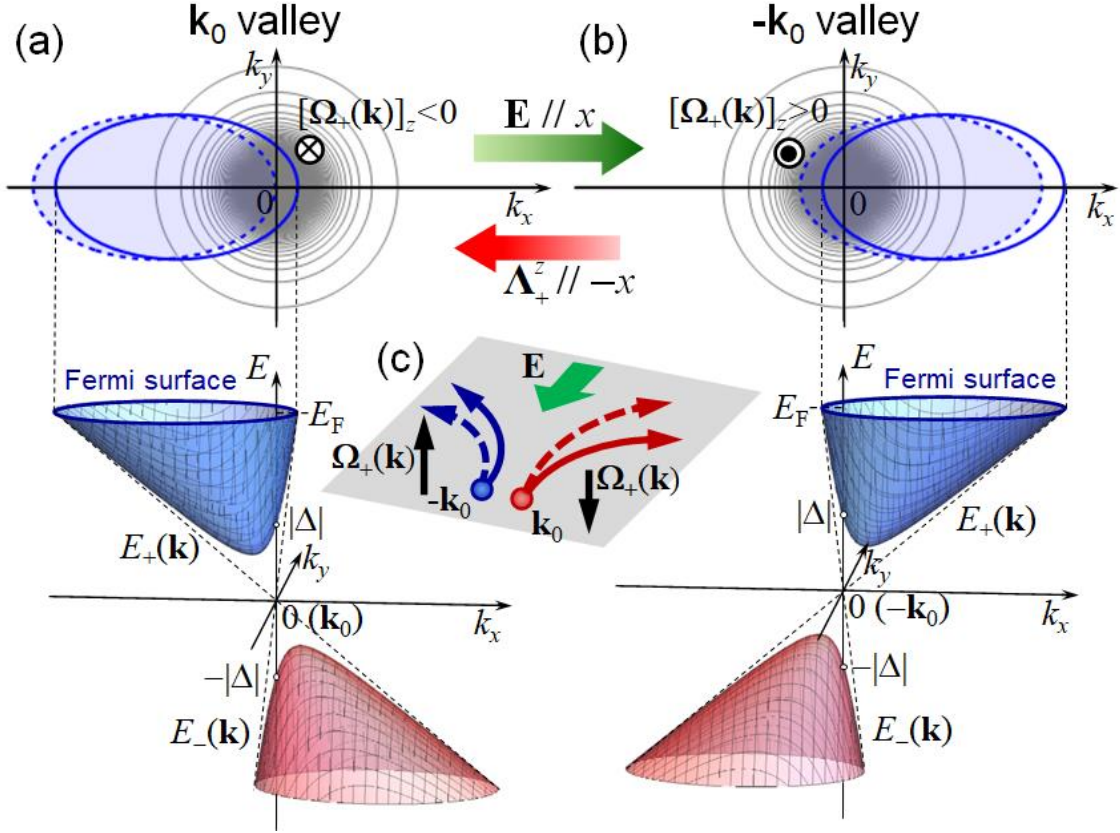

**Figure 1** (color online)

Schematic of the current carrying state in the 2D Dirac fermion system with Dirac cone tilting and gap opening due to inversion symmetry breaking. Here, (a) and (b) represent the electronic structures of the $\mathbf{k}_0$ and $-\mathbf{k}_0$ valleys, respectively. The lower panels of (a) and (b) depict the band dispersion around the energy gap. The upper panels illustrate the Berry curvature (contours) and the Fermi surface of the conduction band. The hatched region indicates the occupied states of the current carrying state under the electric field $\mathbf{E}//x$. $\mathbf{\Lambda}_+^z$ shows the direction of the BCD. (c) Schematic of the mechanism of nonlinear AHE. The balance of average electron motion in the $\mathbf{k}_0$ and $-\mathbf{k}_0$ valleys is broken (dashed arrows) in the current carrying state.

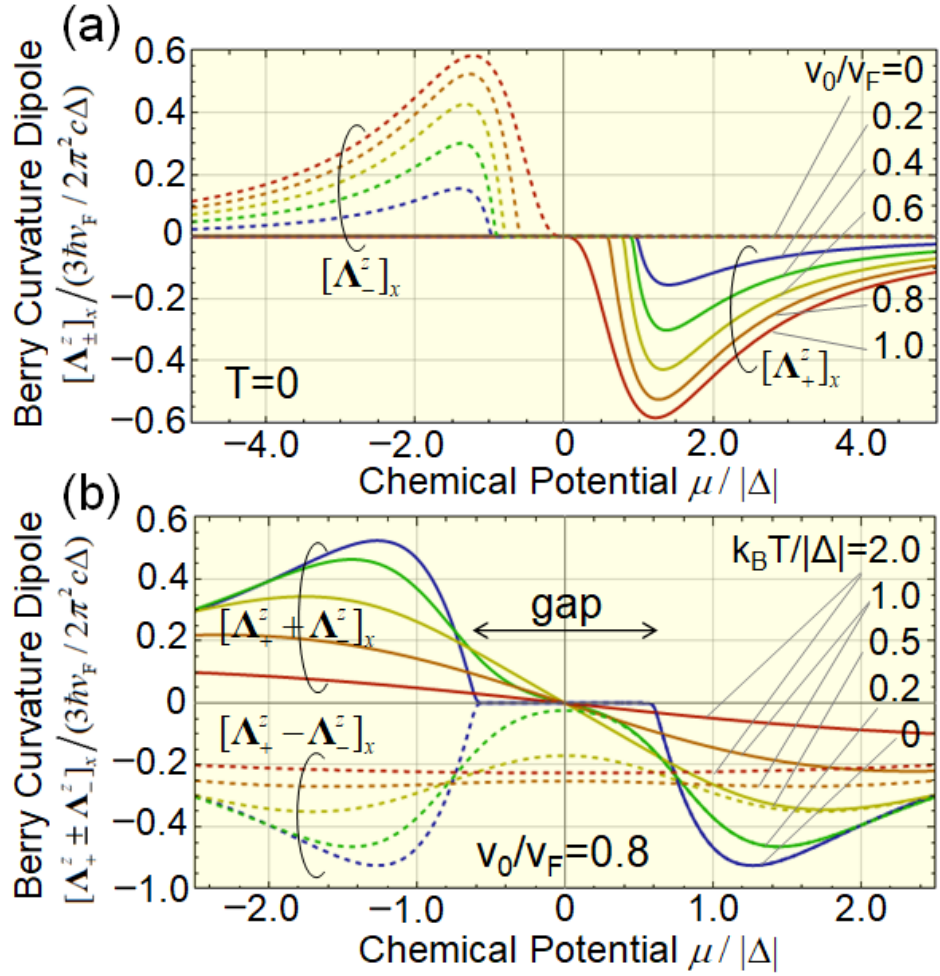

**Figure 2** (color online)

(a) BCD of the conduction (solid lines) and the valence (dashed lines) bands as functions of the chemical potential at $T=0$ for several Dirac cone tilting. (b) Sum (solid lines) and difference (dashed lines) of the BCD of the conduction and the valence bands as functions of the chemical potential at several temperatures for the Dirac cone tilting $v_0/v_F=0.8$.

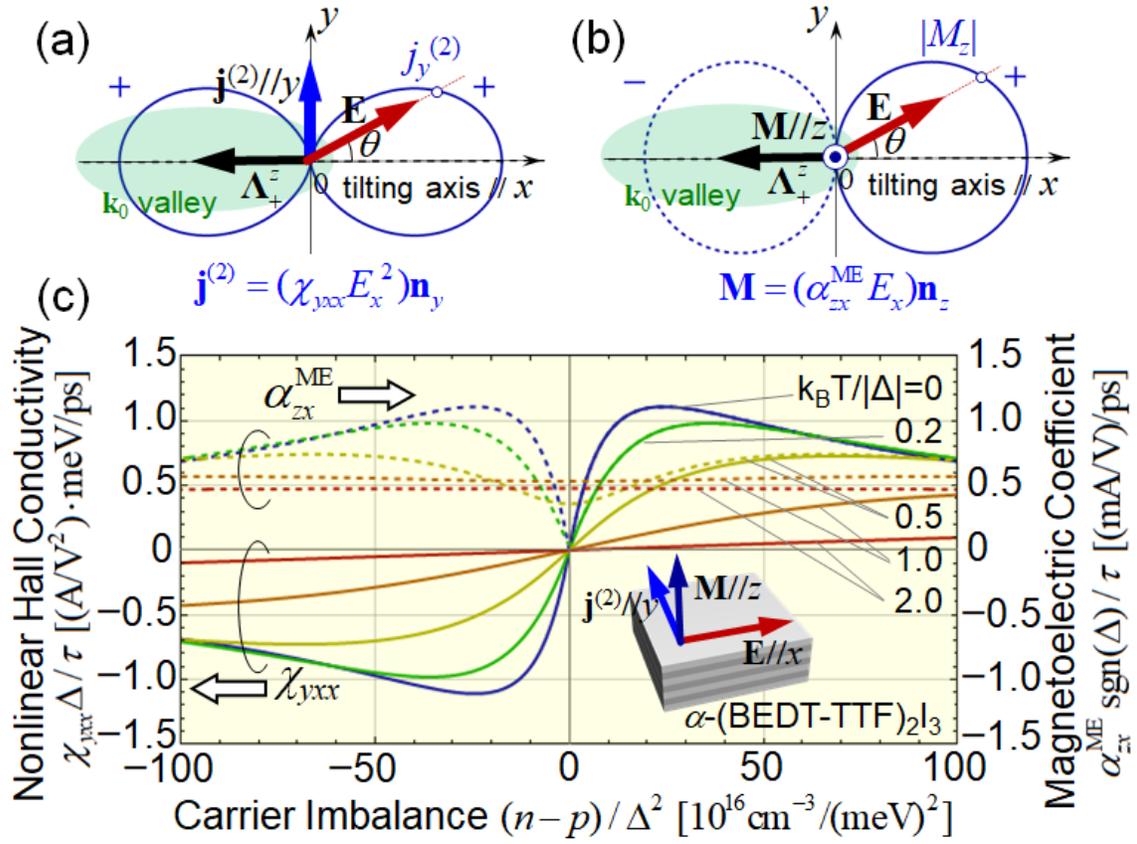

**Figure 3** (color online)

(a) Dependence of the nonlinear anomalous Hall current $j_y^{(2)}$ on the in-plane electric field direction. (b) Dependence of the current-induced orbital magnetization $M_z$ on the in-plane electric field direction. (c) Response coefficients $\chi_{yxx} = j_y^{(2)}/E_x^2$ and $\alpha_{zx}^{ME} = M_z/E_x$ as functions of the difference between the electron and the hole densities $n-p$, estimated using parameters of α-(BEDT-TTF)$_2$I$_3$ ($c = 1.75$ nm, $v_F = 1.0 \times 10^5$ m/s, and $v_0 = 0.8 \times 10^5$ m/s).

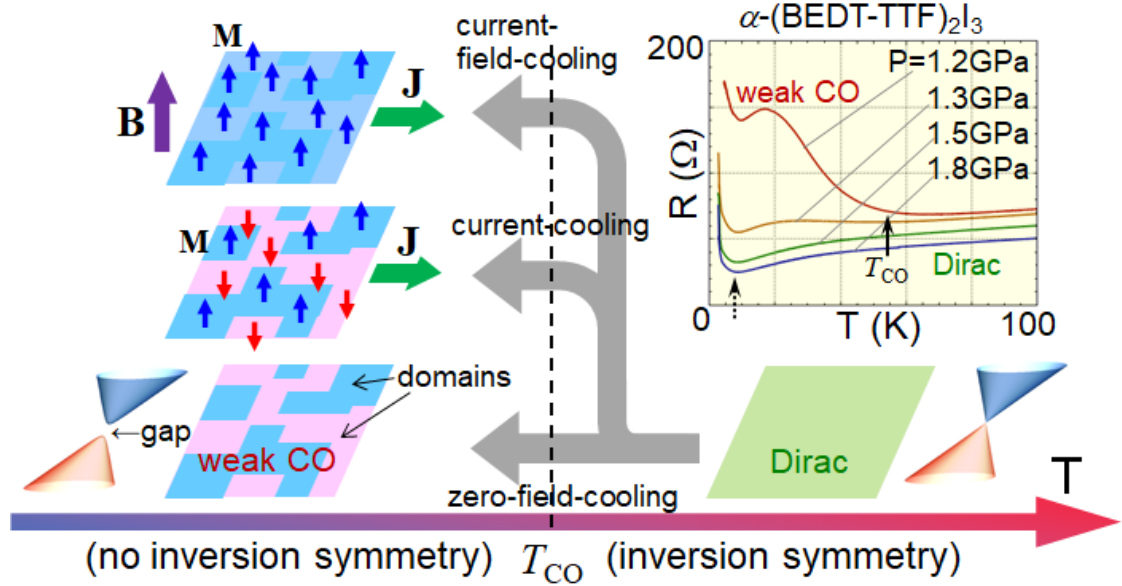

**Figure 4** (color online)

Conceptual representation of the current-field-cooling method. As the system is cooled, the metallic Dirac state undergoes a transition into the weak CO state with multiple inversion asymmetry domains. The electric current induces orbital magnetization in each domain when the current is oriented in the tilting direction. Sample cooling with a finite DC current and an external normal magnetic field is expected to realize the selective formation of one type of domain. Inset: Temperature dependence of interlayer resistance in α-(BEDT-TTF)$_2$I$_3$ after Mori *et al*. [21]. The solid and dotted arrows indicate transitions from the Dirac state into the weak CO state and the unidentified insulating state, respectively.

# Erratum: Possible Current-Induced Phenomena and Domain Control in an Organic Dirac Fermion System with Weak Charge Ordering

[J. Phys. Soc. Jpn. **89**, 103701 (2020)]

Toshihito Osada* and Andhika Kiswandhi

*Institute for Solid State Physics, University of Tokyo,*

*5-1-5 Kashiwanoha, Kashiwa, Chiba 277-8581, Japan.*

In our previous paper [1], we dropped an additional term $\left(\chi_{xxy} E_x E_y\right)\mathbf{n}_x$ of the nonlinear current $\mathbf{j}^{(2)}$. Equation (5) in the paper [1] must be replaced by

$$\begin{aligned}\mathbf{j}^{(2)} &= \left(\chi_{xxy} E_x E_y\right)\mathbf{n}_x + \left(\chi_{yxx} E_x^{\,2}\right)\mathbf{n}_y \\ &= \left\{-(e^3\tau/\hbar^2)[\Lambda_+^z + \Lambda_-^z]_x\right\}|\mathbf{E}|^2 (-\cos\theta\sin\theta\,\mathbf{n}_x + \cos^2\theta\,\mathbf{n}_y),\end{aligned} \quad (1)$$

where $\mathbf{n}_x$ and $\mathbf{n}_y$ are unit vectors in the *x*- and *y*-directions. The nonlinear current is always perpendicular to the electric field **E** with no dissipation. Therefore, this gives the nonlinear anomalous Hall current. The nonlinear current (1) never changes when the electric field **E** is reversed, so that the system shows the rectification characteristics. Correspondingly, Fig. 3 in the previous paper [1] must be replaced by the Fig. 1 in this erratum.

---

*E-mail: osada@issp.u-tokyo.ac.jp

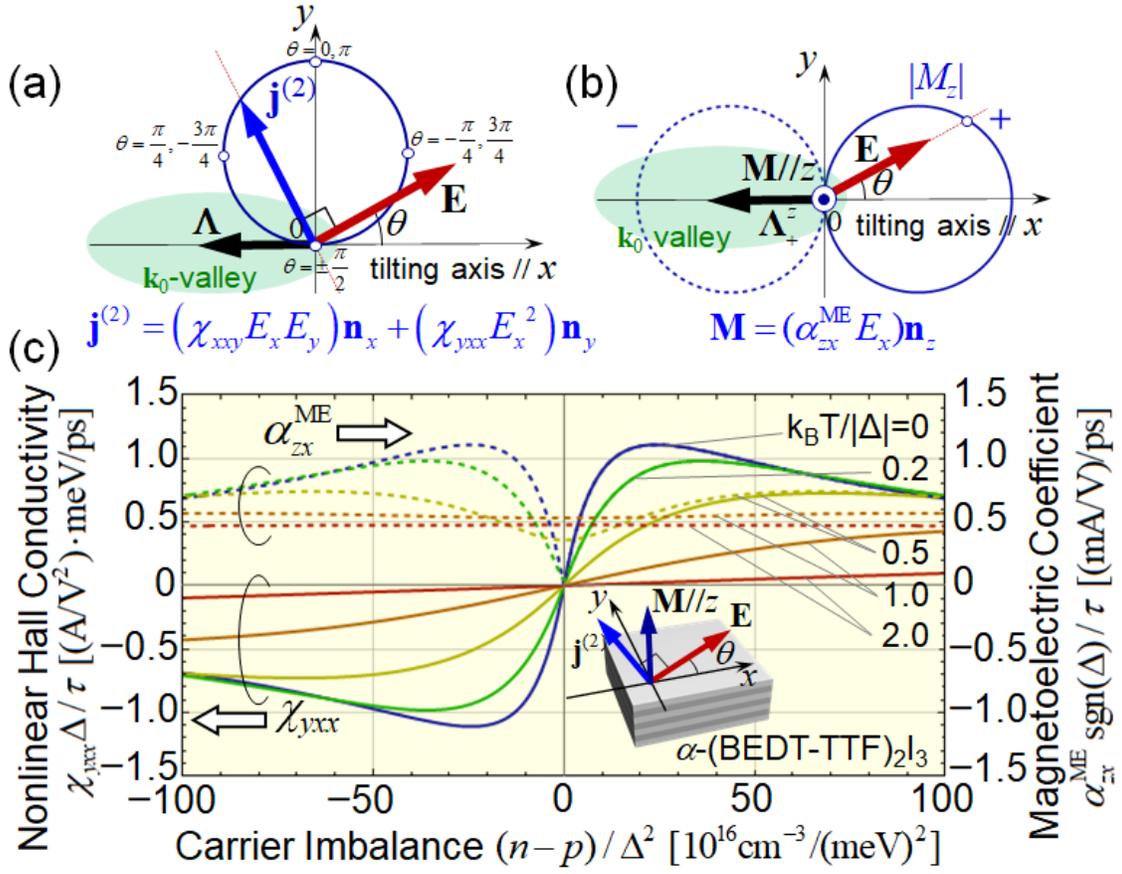

**Figure 1** (color online)

(a) Trajectory of the nonlinear current vector $\mathbf{j}^{(2)}$ when the direction $\theta$ of the in-plane electric field $\mathbf{E}$ is changed. $\chi_{xxy} = -\chi_{yxx}$ is satisfied. (b) Polar plot of the current-induced orbital magnetization $M_z$ as a function of the electric field direction $\theta$. The dashed line indicates negative $M_z$. (c) Response coefficients $\chi_{yxx} = j_y^{(2)}/E_x^2$ and $\alpha_{zx}^{ME} = M_z/E_x$ as functions of the difference between the electron and the hole densities $n-p$, estimated using parameters of α-(BEDT-TTF)$_2$I$_3$ ( $c = 1.75$ nm , $v_F = 1.0 \times 10^5$ m/s , and $v_0 = 0.8 \times 10^5$ m/s ).

18